\documentclass[11pt]{article}

\usepackage[preprint]{acl}

\usepackage{times}
\usepackage{latexsym}

\usepackage[T1]{fontenc}
\usepackage[utf8]{inputenc}

\usepackage{microtype}

\usepackage{inconsolata}

\usepackage{graphicx}

\title{Instructions for *ACL Proceedings}

\author{First Author \\
  Affiliation / Address line 1 \\
  Affiliation / Address line 2 \\
  Affiliation / Address line 3 \\
  \texttt{email@domain} \\\And
  Second Author \\
  Affiliation / Address line 1 \\
  Affiliation / Address line 2 \\
  Affiliation / Address line 3 \\
  \texttt{email@domain} \\}

\usepackage[export]{adjustbox}
\usepackage[ruled]{algorithm2e}
\usepackage[inline, shortlabels]{enumitem}
\usepackage{tcolorbox}
\usepackage{pifont}
\usepackage{xcolor}
\usepackage{xurl}
\usepackage{graphicx}
\usepackage{booktabs}
\usepackage{tabularray}
\usepackage{listings}
\usepackage{amsmath, amsfonts}
\usepackage{nicefrac}
\usepackage{marvosym}
\usepackage{subcaption}
\usepackage{multirow}
\usepackage{makecell}
\usepackage{xspace}
\usepackage{url}
\usepackage{fontawesome5}

\newcommand{\benchmark}{\textsc{STEB}\xspace}
\newcommand{\eg}{e.g.,\xspace}

\title{\textsc{STEB}: A Speech-to-Speech Translation Expressiveness Benchmark\\ for Evaluating Beyond Translation Fidelity}

\author{
  \textbf{Sitong Cheng\textsuperscript{1,2}},
  \textbf{Weizhen Bian\textsuperscript{1}},
  \textbf{Songjun Cao\textsuperscript{2}},
  \textbf{Jin Li\textsuperscript{2}},
  \textbf{Bei Liu\textsuperscript{1}},
  \textbf{Chunyang Jiang\textsuperscript{1}},
\\
  \textbf{Yike Zhang\textsuperscript{2}},
  \textbf{Weihao Wu\textsuperscript{2,3}},
  \textbf{Yiming Li\textsuperscript{1}},
  \textbf{Chi-Min Chan\textsuperscript{1}},
  \textbf{Long Ma\textsuperscript{2}},
  \textbf{Wei Xue\textsuperscript{1}\thanks{Corresponding author.}}
\\
\\
  \textsuperscript{1}Hong Kong University of Science and Technology \\
  \textsuperscript{2}Tencent Youtu Lab \\
  \textsuperscript{3}Shenzhen International Graduate School, Tsinghua University
\\
\small
\faGlobe\ \href{https://cmots.github.io/steb.github.io/}{Homepage}
\hspace{1.2em}
\faGithub\ \href{https://github.com/cmots/STEB}{Code}
}

\begin{document}

\maketitle

\begin{abstract}
Speech-to-speech translation (S2ST) should preserve not only lexical meaning, but also expressive attributes: emotion, scenario style (e.g., news reporting vs.\ dramatic dialogue), and nonverbal vocalizations (NVs). Moreover, collecting cross-lingual target speech that is both translation-faithful and expressively aligned with the source is difficult at scale, making reference-based evaluation impractical.
We introduce \textsc{STEB} (\textbf{S}peech-to-Speech \textbf{T}ranslation \textbf{E}xpressiveness \textbf{B}enchmark), a 32.6-hour Chinese--English benchmark that evaluates both standard dimensions (translation fidelity, speaker similarity, duration alignment) and expressiveness dimensions (emotion, scenario style, NV preservation).
For expressiveness evaluation, \textsc{STEB} uses a caption-then-summarize framework that converts speech into structured expressive attributes and compares source and hypothesis attributes with an LLM judge.
Human validation shows statistically significant correlations with listener judgments across all expressive dimensions.
We evaluate six S2ST systems covering cascaded systems, end-to-end models, and speech large language models. Many systems, especially cascaded ones, achieve strong translation fidelity, but they still struggle with emotion preservation (best: 3.82/5) and NV preservation (best: 2.31/5). 
These results reveal a gap between semantic transfer and expressive transfer, identifying expressiveness preservation as an open challenge for S2ST. Audio samples are available at 
\url{https://cmots.github.io/steb.github.io/}.

\end{abstract}

\section{Introduction}
\label{sec:introduction}

Expressive speech-to-speech translation (S2ST) aims to translate speech across languages while preserving how the source utterance is spoken. This capability is important for real-world applications such as cross-lingual video dubbing, where translation fidelity is insufficient. Recent systems have therefore begun targeting attributes such as timbre preservation, duration alignment, and prosody consistency~\citep{le2024transvip,cheng2025unissunifiedexpressivespeechtospeech,barrault2023seamless}. Yet expressiveness in real-world speech extends beyond timbre or prosody. Three aspects are important for preserving how an utterance is perceived: \textbf{emotion}, \textbf{scenario style} such as news reporting, dramatic dialogue, or audiobook narration, and \textbf{nonverbal vocalizations (NVs)} such as laughter, crying, or breathing.

However, existing benchmarks provide only partial coverage of these expressive dimensions. CVSS~\citep{jia2022cvss} and FLEURS~\citep{conneau2022fleurs} evaluate only translation accuracy without expressive annotations. mExpresso~\citep{barrault2023seamless} and MELD-ST~\citep{chen2024meldst} introduce emotion annotations, but leave scenario style and NV preservation unexamined.

\begin{figure}[t]
    \centering
    \includegraphics[width=\linewidth]{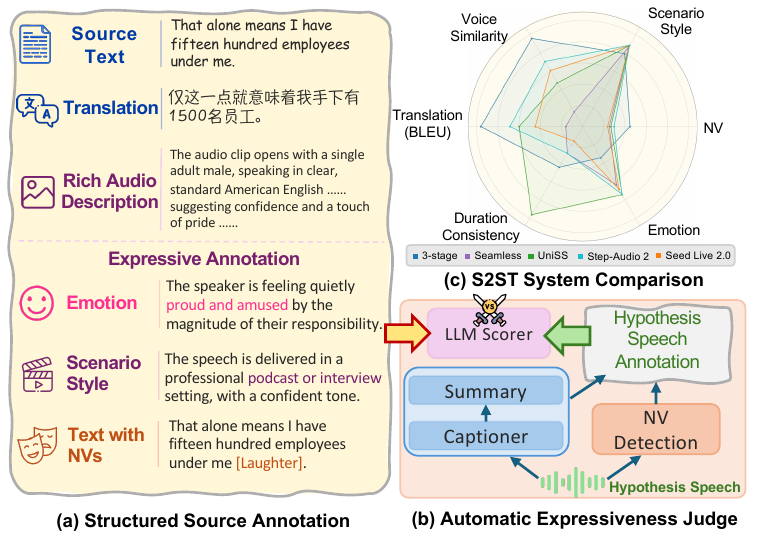}
    \vspace{-2mm}
    \caption{Overview of \benchmark. \textbf{(a)}~Benchmark example showing annotation dimensions: transcription, translation, rich audio description, emotion, scenario style, and text with NVs. \textbf{(b)}~The automatic expressiveness judge: A Captioner--Summary pipeline and NV detection extract hypothesis annotations, which an LLM Scorer compares against source annotations.
    \textbf{(c)}~Radar chart comparing baseline systems across six dimensions.}
    \label{fig:overview}
    \vspace{-2mm}
\end{figure}

Building such a benchmark poses another challenge: expressive S2ST evaluation cannot rely on conventional target-speech references. Constructing parallel S2ST data with expressive consistency across languages is costly and difficult to scale.
Text-to-speech (TTS) synthesis provides limited control over fine-grained expressive attributes. Recruiting human speakers for cross-lingual dubbing with faithful expressiveness reproduction is costly and difficult to scale with diversity. Evaluating expressiveness preservation therefore demands a \emph{reference-free} approach, yet no prior work has established or validated such a framework for S2ST.

To address these gaps, we introduce \benchmark (\textbf{S}peech-to-Speech \textbf{T}ranslation \textbf{E}xpressiveness \textbf{B}enchmark), a benchmark for evaluating expressiveness preservation in S2ST. To the best of our knowledge, \benchmark is the first benchmark that jointly evaluates emotion, scenario style, and NV preservation in S2ST. \benchmark comprises 32.6 hours of Chinese--English evaluation data from six real-world scenarios (drama, audiobooks, advertisements, interviews, news, and commentary). As illustrated in Figure~\ref{fig:overview}, each utterance is annotated with transcription, translation, emotion, scenario style, and rich audio description. Additionally, \benchmark includes an NV subset with 901 utterances and 1.26 hours of speech with confirmed NV annotations.
\benchmark evaluates both standard dimensions (translation fidelity, speaker similarity, and duration alignment) and expressiveness dimensions (emotion, scenario style, and NV preservation). 
For expressiveness evaluation, \benchmark uses a reference-free LLM-as-a-judge framework. The framework first extracts expressive attributes from source and translated speech, then compares them with dimension-specific rubrics on a 1--5 scale. Human correlation studies show statistically significant correlation between automatic scores and human judgments across all expressive dimensions.

We evaluate six S2ST systems on \benchmark, covering cascaded systems, end-to-end S2ST models, and speech large language models. The results show that translation accuracy is relatively strong for many systems, especially cascaded systems. However, expressiveness preservation still lags behind. Current systems achieve much lower scores on emotion consistency and NV preservation than on scenario style consistency. NV preservation benefits from explicit event representation, but even the best system remains far from perfect. 

Our main contributions are as follows:

\begin{itemize}[leftmargin=*, itemsep=2pt, topsep=2pt]
    \item We introduce \benchmark, a 32.6-hour Chinese--English S2ST benchmark covering six real-world scenarios and expressive dimensions including emotion, scenario style, and NVs.
    \item We build a scalable curation pipeline that converts real-world audio into structured source annotations with automatic quality control and human validation.
    \item We propose a reference-free LLM-as-a-judge framework that scores emotion, scenario style, and NV preservation, showing significant correlation with human judgments.
    \item We evaluate six S2ST systems on \benchmark and find that current systems achieve relatively strong translation fidelity but still struggle to preserve emotion and NVs. 
\end{itemize}

\section{Related Work}
\label{sec:related_work}

\paragraph{Speech-to-Speech Translation.}
S2ST has evolved from cascaded automatic speech recognition (ASR), machine translation (MT), and TTS pipelines~\citep{nakamura2006atr, casacuberta2004} through direct spectrogram-based models~\citep{jia2019directs2s, jia2022translatotron2} to discrete unit-based methods~\citep{lee2022directs2sdiscrete, inaguma2023unity, duquenne2023speechmatrix, hwang2024textless}.
More recently, several systems have moved beyond translation fidelity toward expressiveness preservation~\citep{peng2024mslms2st, song2023styles2st}. 
SeamlessExpressive~\citep{barrault2023seamless} preserves prosody and speech rate through a prosody-aware encoder and decoder. 
TransVIP~\citep{le2024transvip} targets voice and isochrony preservation for dubbing. Hibiki~\citep{labiausse2025highfidelitys2sthibiki} addresses expressive S2ST with multi-stream decoding. UniSS~\citep{cheng2025unissunifiedexpressivespeechtospeech} unifies voice, emotion, and duration preservation in a single-stage LLM framework. 

\paragraph{S2ST Benchmarks and Evaluation.} 
As summarized in Table~\ref{tab:benchmark_comparison}, existing benchmarks cover either broad multilingual translation fidelity or isolated expressive dimensions. Durations are reported for each benchmark's test set; for CVSS, FLEURS, and mExpresso, we use the English--Chinese subset. CVSS~\citep{jia2022cvss} synthesizes target speech via TTS, which cannot reproduce fine-grained expressiveness. 
FLEURS~\citep{conneau2022fleurs} offers natural recordings in 102 languages but contains only neutral read-aloud speech without expressiveness annotations. 
Among partially expressive resources, mExpresso~\citep{barrault2023seamless} annotates six read styles with emotion and prosody, and EmphAssess~\citep{deseyssel2024emphassess} evaluates emphasis transfer, a partial form of expressive-style preservation. MELD-ST~\citep{chen2024meldst} annotates emotion for speech translation but does not provide an expressiveness evaluation method. In contrast, \benchmark is the first S2ST benchmark that jointly evaluates emotion, scenario style, and NV preservation.

\begin{table}[t]
    \centering
    \caption{Comparison of S2ST evaluation benchmarks. \textbf{Hrs} denotes test-audio duration. Dimensions: emotion preservation (\textbf{Emo.}), scenario-style preservation (\textbf{Sty.}), and nonverbal-vocalization preservation (\textbf{NV}). \ding{51}: supported; \ding{55}: not supported; 
\ding{51}\rotatebox[origin=c]{-9.2}{\kern-0.7em\ding{55}}: partial.}
    \label{tab:benchmark_comparison}
    \vspace{-2mm}
    \resizebox{\linewidth}{!}{
    \begin{tabular}{l c c  c c c}
        \toprule
        \textbf{Benchmark} & \textbf{Domains} & \textbf{Hrs} & \textbf{Emo.} & \textbf{Sty.} & \textbf{NV}  \\
        \midrule
        CVSS & Open domain & 12.1 & \ding{55} & \ding{55} & \ding{55} \\
        FLEURS & Wikipedia & 4.8  & \ding{55} & \ding{55} & \ding{55} \\
        mExpresso & Read & 4.8  & \ding{51} & \ding{55} & \ding{55} \\
        MELD-ST & Drama & 3.3  & \ding{51} & \ding{55} & \ding{55} \\
        EmphAssess & Read & 2.1 & \ding{55} & \ding{51}\rotatebox[origin=c]{-9.2}{\kern-0.7em\ding{55}} & \ding{55} \\
        \midrule
        \benchmark (Ours) & Open domain & 32.6  & \ding{51} & \ding{51} & \ding{51} \\
        \bottomrule
    \end{tabular}
    }
    \vspace{-2mm}
\end{table}

\paragraph{LLM-based Evaluation.}
Large language models are increasingly used as automated evaluators for speech tasks.  Audio-language benchmarks including AIR-Bench~\citep{yang2024airbenchbenchmarkinglargealaudiolanguagemodels} and AudioBench~\citep{wang2025audiobench} confirm that audio LLMs can reason about paralinguistic properties. \citet{chiang2025audioawarelargelanguagemodelsjudges} show that audio-aware LLMs can judge generated speech on emotion, volume, and speaking pace, achieving agreement with human raters comparable to inter-annotator agreement. TTS-PRISM~\citep{wang2026ttsprismperceptualreasoninginterpretable} and EchoMind~\citep{zhou2026echomindinterrelatedmultilevelbenchmark} employ LLMs for multi-dimensional speech evaluation covering emotion, style, and paralinguistic attributes. However, no prior work has validated an LLM-based framework specifically for evaluating expressiveness preservation in S2ST.

\section{Methods}
\label{sec:methods}

\begin{figure*}[t]
    \centering
    \includegraphics[width=\linewidth]{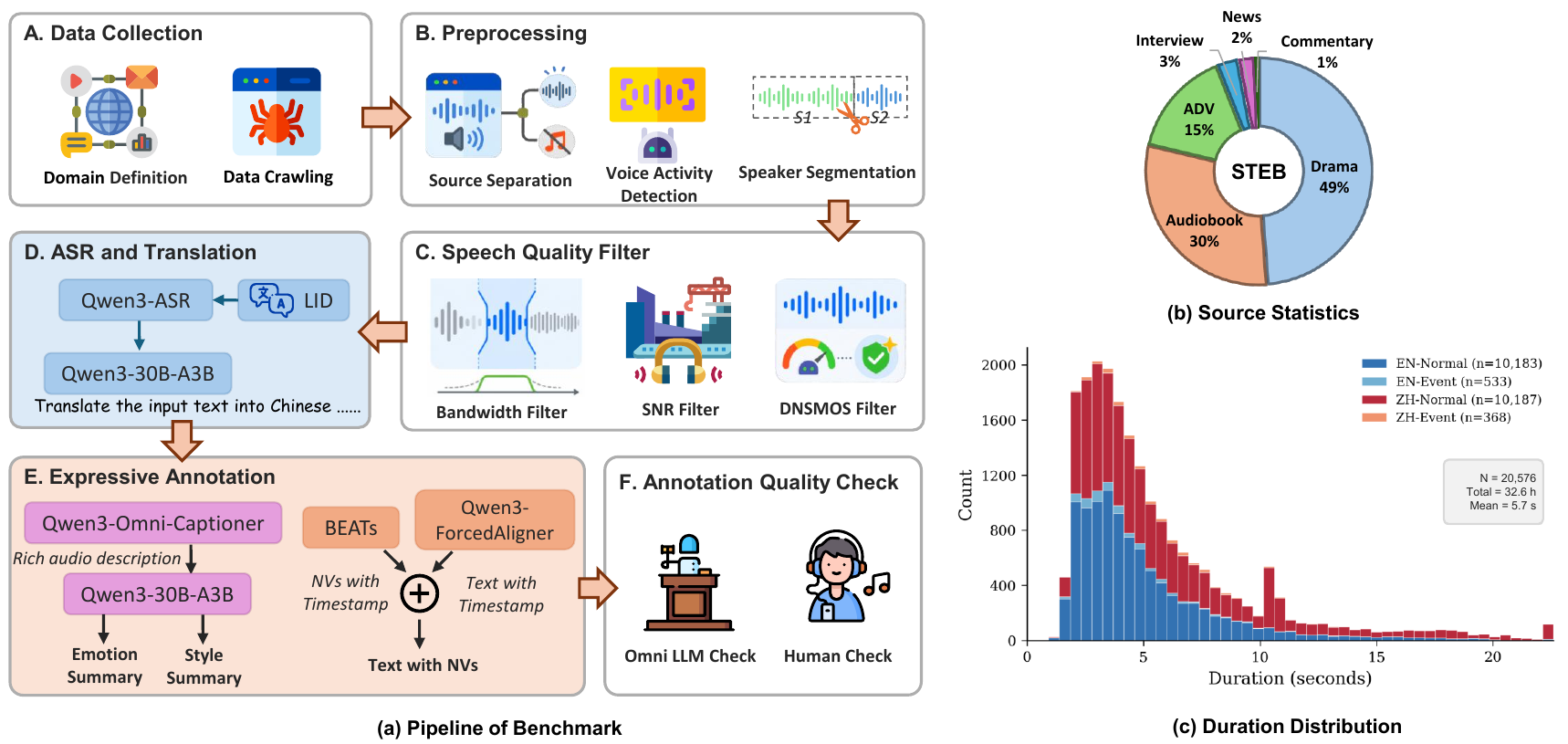}
    \vspace{-2mm}
    \caption{Overview of the \benchmark data curation pipeline. The pipeline consists of six stages: \textbf{(A)} data collection from six real-world scenarios; \textbf{(B)} preprocessing for clean single-speaker utterance extraction; \textbf{(C)} speech quality filtering; \textbf{(D)} transcription, language identification, and translation; \textbf{(E)} expressive annotation for emotion, scenario style, and NVs; and \textbf{(F)} annotation quality checking. The right panels show the scenario-source distribution and utterance-duration distribution of the final benchmark.}
    \label{fig:pipeline}
    \vspace{-2mm}
\end{figure*}

Figure~\ref{fig:pipeline} illustrates the \benchmark construction pipeline, which comprises data collection and preprocessing, expressiveness annotation, and quality assurance. We also describe the automatic expressiveness judge method for S2ST in Section~\ref{ssec:judge}.

\subsection{Data Collection and Preprocessing}
\label{ssec:collection}

We collect candidate audio from publicly accessible web sources across six real-world scenarios: \textbf{drama}, \textbf{audiobooks}, \textbf{advertisements}, \textbf{interviews}, \textbf{news broadcasts}, and \textbf{commentary}. These recordings span minutes to hours, mix background music with speech, and involve multiple speakers. Direct utterance-level evaluation on such raw audio is unreliable. Therefore, we apply the following stages to extract clean single-speaker utterances:
\begin{enumerate}[leftmargin=*, itemsep=1pt, topsep=2pt]
    \item \textbf{Source separation}: We separate background music from foreground speech using BS-Roformer~\citep{lu2024music}. This reduces non-speech interference and improves the robustness of downstream annotation.
    \item \textbf{Speaker segmentation}: We first detect speech activity with Silero VAD~\citep{silerovad} to locate voiced regions, then perform speaker diarization with pyannote~\citep{bredin2020pyannoteaudio} to identify turn boundaries. Same-speaker segments are merged using CAM++~\citep{wang2023campp} speaker embeddings (cosine similarity threshold $\geq 0.75$). This procedure produces single-speaker utterances suitable for per-speaker expressiveness annotation.
    \item \textbf{Quality filtering}: Segments outside the 3--30 second range are discarded---fragments shorter than 3 seconds lack sufficient context for meaningful expressiveness annotation, while segments longer than 30 seconds are too long for reliable utterance-level annotation.
    We then apply three sequential quality checks: (1)~a narrowband filter removes recordings with limited bandwidth, (2)~a signal-to-noise ratio (SNR) check discards segments with excessive background noise, and (3)~DNSMOS~\citep{reddy2021dnsmos} perceptual quality scoring removes segments below a threshold of 3.0.
    \item \textbf{Language identification}: We retain only Chinese and English utterances using Whisper~\citep{radford2023whisper} language detection. Segments in other languages or with low confidence scores are discarded.
\end{enumerate}
More details of preprocessing pipeline are listed in Appendix~\ref{app:preprocessing}.

\subsection{Annotation}
\label{ssec:annotation}

Each \benchmark dimension requires structured annotations on the source utterance to serve as reference annotations. We employ an automatic pipeline combining multimodal LLMs with task-specific models to produce these annotations at scale, reducing the need for manual annotation.

\paragraph{ASR, force alignment and translation}
All retained segments are transcribed using Qwen3-ASR~\citep{shi2026qwen3asrtechnicalreport}, and the resulting source transcriptions are used as input for translation. We further apply Qwen3-ForceAlign~\citep{shi2026qwen3asrtechnicalreport} to obtain word-level timestamps for downstream NVs alignment.
Source transcriptions are then translated into the target language (Chinese$\leftrightarrow$English) with Qwen3-30B-A3B~\citep{yang2025qwen3technicalreport}. 

\paragraph{NV detection.}
NVs are detected using BEATs~\citep{chen2023beats} within the PretrainedSED framework~\citep{schmid2025effectivepretrainingaudiotransformers}. NV markers are inserted into the transcription according to their temporal overlap with adjacent word intervals. The result is transcriptions with NV markers where NVs appear as bracketed markers within the transcription (\eg ``\texttt{[Breathing] That's amazing [Laughter].}'').

\paragraph{Audio captioning and summarization.}
The annotation pipeline follows a two-stage \emph{caption-then-summarize} design. In the first stage, Qwen3-Omni-Captioner~\citep{xu2025qwen3omnitechnicalreport} generates a detailed audio caption for each utterance, covering vocal emotion, scene/genre, and delivery cues (pace, energy, formality, timbre). The captioner operates directly on the audio signal, capturing paralinguistic information that text-only analysis would miss. In the second stage, a text LLM (Qwen3-30B-A3B) summarizes the verbose caption into an \emph{emotion} and a \emph{scenario style} description separately.

\subsection{Quality Assurance and Dataset Statistics}
\label{ssec:quality}

The reliability of \benchmark depends on the accuracy of its annotations. We therefore use both automatic quality scoring and human validation. Starting from 80 hours of candidate data selected from a large raw pool, the final benchmark retains fewer than 40 hours, with a selection rate below 50\%.

\paragraph{Rule-based filtering.} Candidate utterances are selected using rules that prioritize those containing NVs, while improving coverage across scenarios, languages (Chinese and English), and speaker genders. A text LLM is also used as a secondary language check based on transcription. It catches errors missed by Whisper and removes samples with heavy Chinese--English code-mixing, which would confound monolingual evaluation.

\paragraph{Automatic quality scoring.} We employ a strong multimodal LLM as an automatic quality judge to verify the accuracy of the generated annotations. The model is provided with each audio sample and its text annotation, including the transcription, translation, emotion, scenario style, and transcription with NV. It then scores the relevant quality dimensions on a 1--5 scale. The scoring prompt is in Appendix~\ref{app:gemini_prompt}.

We apply dimension-specific retention thresholds with a target of perfect scores. For the normal subset, we evaluate four dimensions: transcription, translation, emotion, and scenario style. All dimensions must score 5, with at most one dimension allowed to score 4. For the NV subset, we evaluate three dimensions: transcription, translation, and transcriptions with NV markers.

This strict criterion acts as an automatic quality gate. Among annotated candidates, 88.3\% receive an emotion score of 5, and 91.4\% receive a scenario style score of 5. We also compare the automatic scores with human ratings on a random subset of 170 samples. When the automatic score is above 4, the human score is also above 4 in 100\% of cases for transcription, 91.3\% for translation, 86.7\% for emotion, 88.8\% for scenario style, and 75.6\% for NV annotation. These results suggest that the automatic quality gate is reliable for most dimensions.

\paragraph{Human validation.} 
We further conduct human validation to check the quality of retained utterances in normal and NV subsets after automatic quality filtering. Annotators review both speech quality and annotation accuracy. Speech-quality checks includes multi-speaker contamination, incorrect language labels, and low-quality recordings. Annotation accuracy covers transcription, translation, emotion, scenario style, and NV annotations. Details of the human evaluation criteria are provided in Appendix~\ref{app:human_eval}.

\paragraph{Dataset statistics.} The final benchmark comprises two subsets: a \emph{normal subset} (20,370 utterances, 32.27 hours) covering standard speech across all six scenarios, and an \emph{NV subset} (901 utterances, 1.26 hours) containing confirmed NV annotations. Some utterances in the NV subset are also included in the normal subset, so the total unique duration of the benchmark is 32.6 hours. Figure~\ref{fig:pipeline} (b) and (c) shows the scenario source distribution and utterance duration distribution. Drama accounts for the largest share because it provides dense emotional variation. The other scenarios improve coverage of scenario styles. The duration range of 3--30 seconds covers both short conversational turns and longer narrative passages.

\subsection{Reference-Free Expressiveness Scoring}
\label{ssec:judge}

Evaluating expressiveness preservation in S2ST requires a reference-free approach because ground-truth target speech that faithfully preserves source expressiveness is not available at scale. A possible alternative is to classify the source and target independently and compare the predicted labels. However, this approach is insufficient. Label matching cannot capture within-category quality differences, such as emotion intensity. It also cannot handle NV preservation, where type, count, and position must be compared at the event level. \benchmark therefore formulates expressiveness evaluation as a graded comparative task. It adopts an LLM-as-a-judge approach to compare source and translated speech through structured textual descriptions, without requiring parallel audio references.

\paragraph{Judge pipeline.}
Given source speech $X$ with expressiveness annotations (emotion $E_x$, scenario style $\mathit{ST}_x$, transcriptions with NV markers $N_x$) and a system's translated output $Y$, the judge operates in two stages. First, $Y$ is processed through the same captioning-then-summarization pipeline used for source annotation (Section~\ref{ssec:annotation}): Qwen3-Omni-Captioner generates an audio caption $C_y$ of $Y$, and Qwen3-30B-A3B summarizes $C_y$ into hypothesis emotion $E_y$ and scenario style $\mathit{ST}_y$. BEATs detects NVs in $Y$ and aligns them to produce $N_y$. Second, an LLM judge scores the consistency between source and hypothesis descriptions on a 1--5 scale with dimension-specific rubrics.

\paragraph{Scoring rubrics.}
Each expressiveness dimension is scored on a 1--5 scale with a structured rubric:
\begin{itemize}[leftmargin=*, itemsep=1pt, topsep=2pt]
    \item \textbf{Emotion}: Whether the core emotion category, intensity, and attitude are preserved.
    \item \textbf{Scenario style}: Whether the scenario style is consistent; formality and delivery manner serve as secondary cues.
    \item \textbf{NV preservation}: Whether NV type, count, order, and approximate position match between source and hypothesis NV-marked transcriptions.
\end{itemize}
Full rubrics with criteria for each score level are in Appendix~\ref{app:judge_prompts}.

\paragraph{Aggregation.}

To reduce scoring variance, each sample is judged three times independently. We apply a simple outlier-aware aggregation rule to obtain the final score from the three runs; full details are in Appendix~\ref{app:aggregation}.

\section{Experiments}
\label{sec:experiments}

\subsection{Experimental Setup}
\label{ssec:setup}

\paragraph{Evaluated systems.} We evaluate six S2ST systems on \benchmark, covering cascaded systems, end-to-end models, and speech LLMs. Full details are in Appendix~\ref{app:system_details}.
\begin{itemize}[leftmargin=*, itemsep=1pt, topsep=2pt]
    \item \textbf{Three-Stage}: A three-stage cascade comprising ASR with NV detection, text LLM translation with style instruction generation, and instruction-controlled voice-cloning TTS. The pipeline transcribes source speech with Qwen3-ASR, detects NVs, and passes the result to Qwen3-235B-A22B~\citep{yang2025qwen3technicalreport} for translation with emotion and style instruction generation. We select VoxCPM2~\citep{voxcpm2_2026} as the TTS module because it simultaneously supports voice cloning and instruction-controlled generation. We report a \emph{w/o Instruct} variant where TTS receives text and reference audio without style instructions.
    \item \textbf{Two-Stage}: A two-stage cascade using Qwen3-Omni-Instruct for direct audio-to-text translation with emotion and style instruction, followed by VoxCPM2 TTS with voice cloning. We also report a \emph{w/o Instruct} variant.
    \item \textbf{UniSS}~\citep{cheng2025unissunifiedexpressivespeechtospeech}: A unified single-stage expressive S2ST model that handles translation, voice preservation, emotion transfer, and duration alignment within a single framework.
    \item \textbf{SeamlessExpressive}~\citep{barrault2023seamless}: A multilingual end-to-end S2ST model, which adopts a unit-based discrete speech representation with a prosody-aware encoder.
    \item \textbf{Seed LiveInterpret 2.0}~\citep{cheng2025seedliveinterpret20endtoend}: A commercial simultaneous speech translation system with duplex speech understanding and generation capabilities.
    \item \textbf{Step-Audio~2}~\citep{wu2025stepaudio2technicalreport}: An end-to-end speech large language model supporting direct S2ST via large-scale speech--text pretraining. 
\end{itemize}

\paragraph{Metrics.} We report all \benchmark dimensions. For translation accuracy, we use BLEU~\citep{papineni2002bleu}, COMET~\citep{rei2020comet}, and XCOMET~\citep{guerreiro2024xcomet}. We report both reference-based and quality-estimation (QE) variants of XCOMET, where the QE variant requires no reference translation. For expressiveness, we evaluate emotion consistency (Emo.), scenario style consistency (Sty.), and NV preservation (NV.) via LLM-judged 1--5 scores. NV preservation is evaluated only on the NV subset. Speaker similarity (SIM) is measured as cosine similarity from WavLM-based~\citep{Wang2021UniSpeech} speaker embeddings. Duration alignment is quantified by SLC at $\epsilon=0.2$ and $\epsilon=0.4$~\citep{wu2023videodubber}, measuring the proportion of utterances whose duration ratio falls within the tolerance. Detailed metric definitions are provided in Appendix~\ref{app:metrics}.

\subsection{Main Results}
\label{ssec:main_results}

\begin{table*}[t]
    \centering
    \caption{Main comparison results on \benchmark (Chinese$\rightarrow$English). Higher scores indicate better performance. Best results are in \textbf{bold}.}
    \label{tab:main_zh2en}
    \vspace{-2mm}
    \resizebox{\linewidth}{!}{
    \begin{tabular}{l ccc ccc cc c}
        \toprule
        \multirow{2}{*}{\textbf{System}} & \multicolumn{3}{c}{\textbf{Translation}} & \multicolumn{3}{c}{\textbf{Expressiveness}} & \multicolumn{2}{c}{\textbf{Duration}} & \textbf{Voice} \\
        \cmidrule(lr){2-4} \cmidrule(lr){5-7} \cmidrule(lr){8-9} \cmidrule(lr){10-10}
        & BLEU $\uparrow$ & COMET $\uparrow$ & XCOMET/QE $\uparrow$ & Emo. $\uparrow$ & Sty. $\uparrow$ & NV. $\uparrow$ & SLC$_{0.2}$ $\uparrow$ & SLC$_{0.4}$ $\uparrow$ & SIM $\uparrow$ \\
        \midrule
        Three-Stage               & 38.20 & 0.812 & \textbf{0.804}/\textbf{0.840} & 1.73 & 3.92 & 2.25 & 0.614 & 0.887 & 0.497 \\
        \quad w/o Instruct        & 38.21 & 0.812 & \textbf{0.804}/\textbf{0.840} & 1.73 & 3.93 & 2.22 & 0.610 & 0.889 & \textbf{0.498} \\
        Two-Stage                 & \textbf{41.59} & 0.814 & 0.800/0.833 & 1.68 & 3.93 & \textbf{2.31} & 0.439 & 0.699 & \textbf{0.498} \\
        \quad w/o Instruct        & 41.41 & \textbf{0.815} & 0.800/0.834 & 1.67 & 3.92 & 2.28 & 0.438 & 0.696 & 0.497 \\
        \midrule
        UniSS                     & 28.55 & 0.768 & 0.772/0.793 & \textbf{3.61} & 4.36 & 1.31 & \textbf{0.915} & \textbf{0.959} & 0.411 \\
        Seamless                  & 17.83 & 0.685 & 0.711/0.693 & 3.10 & 4.27 & 1.29 & 0.598 & 0.934 & 0.302 \\
        Seed Live 2.0             & 26.02 & 0.763 & 0.773/0.798 & 3.34 & \textbf{4.41} & 1.25 & 0.590 & 0.912 & 0.416 \\
        Step-Audio~2              & 30.82 & 0.784 & 0.790/0.814 & 3.57 & 4.39 & 1.58 & 0.542 & 0.836 & 0.475 \\
        \bottomrule
    \end{tabular}
    }
    \vspace{-2mm}
\end{table*}

\begin{table*}[t]
    \centering
    \caption{Results on \benchmark (English$\rightarrow$Chinese). Same metrics as Table~\ref{tab:main_zh2en}. Best results are in \textbf{bold}.}
    \label{tab:main_en2zh}
    \vspace{-2mm}
    \resizebox{\linewidth}{!}{
    \begin{tabular}{l ccc ccc cc c}
        \toprule
        \multirow{2}{*}{\textbf{System}} & \multicolumn{3}{c}{\textbf{Translation}} & \multicolumn{3}{c}{\textbf{Expressiveness}} & \multicolumn{2}{c}{\textbf{Duration}} & \textbf{Voice} \\
        \cmidrule(lr){2-4} \cmidrule(lr){5-7} \cmidrule(lr){8-9} \cmidrule(lr){10-10}
        & BLEU $\uparrow$ & COMET $\uparrow$ & XCOMET/QE $\uparrow$ & Emo. $\uparrow$ & Sty. $\uparrow$ & NV. $\uparrow$ & SLC$_{0.2}$ $\uparrow$ & SLC$_{0.4}$ $\uparrow$ & SIM $\uparrow$ \\
        \midrule
        Three-Stage               & 54.14 & 0.892 & \textbf{0.849}/0.895 & 1.68 & 4.02 & \textbf{2.21} & 0.659 & 0.917 & \textbf{0.428} \\
        \quad w/o Instruct        & 54.09 & 0.891 & \textbf{0.849}/0.894 & 1.70 & 4.03 & 2.09 & 0.652 & 0.922 & 0.427 \\
        Two-Stage                 & 59.66 & \textbf{0.902} & \textbf{0.849}/\textbf{0.902} & 1.70 & 4.03 & 2.10 & 0.641 & 0.908 & 0.427 \\
        \quad w/o Instruct        & \textbf{59.73} & \textbf{0.902} & \textbf{0.849}/\textbf{0.902} & 1.69 & 4.03 & 2.05 & 0.638 & 0.902 & 0.425 \\
        \midrule
        UniSS                     & 46.43 & 0.849 & 0.824/0.862 & \textbf{3.82} & 4.44 & 1.36 & \textbf{0.980} & \textbf{0.990} & 0.291 \\
        Seamless                  & 36.13 & 0.806 & 0.767/0.794 & 3.27 & 4.39 & 1.23 & 0.492 & 0.819 & 0.254 \\
        Seed Live 2.0             & 41.72 & 0.805 & 0.814/0.854 & 3.53 & 4.41 & 1.09 & 0.342 & 0.819 & 0.348 \\
        Step-Audio~2              & 48.37 & 0.865 & 0.838/0.880 & 3.77 & \textbf{4.46} & 1.38 & 0.548 & 0.874 & 0.334 \\
        \bottomrule
    \end{tabular}
    }
    \vspace{-2mm}
\end{table*}

Tables~\ref{tab:main_zh2en} and~\ref{tab:main_en2zh} present results for Chinese$\rightarrow$English and English$\rightarrow$Chinese respectively. We highlight three findings.

\paragraph{(1) Translation accuracy is relatively strong, but expressiveness lags behind.}
Most systems achieve reasonable translation accuracy across both directions. Cascaded systems obtain the strongest translation scores, with Two-Stage reaching 41.59 BLEU for Chinese$\rightarrow$English and 59.66 BLEU for English$\rightarrow$Chinese. Other models also achieve competitive COMET and XCOMET scores, suggesting that content transfer is not the main failure mode for many current systems.

In contrast, expressiveness preservation remains much weaker. Cascaded systems obtain emotion scores of only 1.67--1.73, while end-to-end systems reach 3.10--3.82. Scenario style follows a similar trend but with a smaller gap, suggesting that it is less degraded than emotion in current systems.

However, adding instructions to cascaded TTS brings almost no gain over the w/o Instruct variants. This suggests that textual instructions are insufficient for recovering source expressiveness in cascaded S2ST systems. Overall, current systems are much better at preserving what is said than how it is said.

\paragraph{(2) NV preservation benefits from explicit representation.}
NV preservation shows a different pattern from emotion and scenario style. Systems that use explicit NV markers obtain higher NV scores than systems that generate speech directly. Two-Stage reaches 2.31 for Chinese$\rightarrow$English, and Three-Stage reaches 2.21 for English$\rightarrow$Chinese. Both systems expose NVs as inline text markers before TTS generation.

In contrast, end-to-end speech systems score no higher than 1.58. Although Step-Audio~2 is prompted to preserve NVs during reasoning, the results suggest that such information is often not carried into the final speech output. Unlike emotion and scenario style, NVs can be represented as discrete events. This makes text-level markers useful, but the low absolute scores show that translation with NVs in speech remains difficult.

\paragraph{(3) Duration alignment needs dedicated duration control.}
UniSS achieves the best duration alignment by a large margin. Its SLC$_{0.2}$ reaches 0.915 for Chinese$\rightarrow$English and 0.980 for English$\rightarrow$Chinese, consistent with its explicit duration-control design. All other systems are much lower. The next best system reaches only 0.614/0.659 on SLC$_{0.2}$. 
This result shows that duration alignment requires explicit control over the length of the generated speech.
Such control is important for use cases such as video dubbing and simultaneous speech translation, where the translated speech must match the timing of the source.

\paragraph{Text-based translation analysis.}
We further evaluate the intermediate text outputs of each system before speech generation. Table~\ref{tab:text_results} reports BLEU, COMET, and XCOMET on these text outputs. The text scores are close to those measured from the final speech outputs. For example, Two-Stage achieves 43.65 BLEU on text and 41.59 BLEU on final speech for Chinese$\rightarrow$English, and 61.51 vs.\ 59.66 for English$\rightarrow$Chinese. Many baselines show similar small gaps.
These results suggest the main bottleneck is not semantic transfer during speech generation, but expressive preservation across the full speech-to-speech pipeline.

\begin{table}[t]
    \centering
    \caption{Text-based translation fidelity of system intermediate outputs. Higher scores indicate better performance. Best results are in \textbf{bold}.}
    \label{tab:text_results}
    \vspace{-2mm}
    \resizebox{\linewidth}{!}{
    \begin{tabular}{l ccc ccc}
        \toprule
        & \multicolumn{3}{c}{\textbf{zh$\rightarrow$en}} & \multicolumn{3}{c}{\textbf{en$\rightarrow$zh}} \\
        \cmidrule(lr){2-4} \cmidrule(lr){5-7}
        \textbf{System} & BLEU & COMET & XCOMET/QE & BLEU & COMET & XCOMET/QE \\
        \midrule
        Three-Stage    & 40.61 & 0.826 & 0.812/\textbf{0.859} & 55.30 & 0.901 & 0.864/0.911 \\
        Two-Stage      & \textbf{43.65} & \textbf{0.834} & \textbf{0.813}/0.857 & \textbf{61.51} & \textbf{0.914} & \textbf{0.867}/\textbf{0.922} \\ \midrule
        UniSS          & 30.87 & 0.791 & 0.793/0.814 & 48.91 & 0.875 & 0.840/0.878 \\
        Seamless       & 17.66 & 0.695 & 0.720/0.702 & 37.87 & 0.824 & 0.805/0.825 \\
        Seed Live 2.0  & 26.20 & 0.765 & 0.774/0.796  & 47.99 & 0.876 & 0.841/0.885 \\
        Step-Audio~2   & 33.68 & 0.802 & 0.803/0.836 & 50.61 & 0.891 & 0.856/0.900 \\
        \bottomrule
    \end{tabular}
    }
    \vspace{-2mm}
\end{table}

\subsection{Human Correlation Study}
\label{ssec:human_correlation}

We validate the \benchmark evaluation framework through a human correlation study on three expressiveness dimensions.

\paragraph{Judge designs.}
For expressiveness evaluation, we compare several judge designs under similar rubrics to understand how different designs align with human judgments. We investigate three methods of scorers: 
\begin{itemize}[leftmargin=*, itemsep=1pt, topsep=2pt]
    \item \textbf{Summary-based scoring}: Uses the captioning-then-summarization pipeline (Section~\ref{ssec:judge}) to extract structured annotations, then Qwen3-30B-A3B scores source--hypothesis consistency with dimension-specific prompts and rubrics.
    \item \textbf{Caption-based scoring}: The scorer receives the full audio captions of the source and hypothesis. It scores each dimension separately using the corresponding prompt and rubric.
    \item \textbf{Audio-based direct scoring}: A multimodal LLM directly listens to both source and hypothesis audio and scores their expressiveness consistency without intermediate text annotations.
\end{itemize}

\paragraph{Setup.}
We evaluate the automatic judge on 160 human-scored source-hypothesis pairs across the three expressiveness dimensions. Each pair is scored by at least three independent annotators using the same 1--5 scale as the automatic framework. To obtain reliable human reference scores for correlation computation, we retain only pairs with sufficient inter-annotator consistency. Details of the human annotation protocol and consistency filtering are provided in Appendix~\ref{app:judge_human_eval}.

\paragraph{Results.}
Table~\ref{tab:human_corr} reports Spearman's rank correlation ($\rho$)~\citep{spearman1904proof}, $p$-values, exact agreement (score difference $\leq 0.5$), and mean absolute error (MAE). We compare human--human agreement (H--H) with three LLM judge designs: summary-based (L\textsubscript{s}--H), direct audio-based (L\textsubscript{d}--H), and caption-based (L\textsubscript{c}--H).

\begin{table}[t]
    \centering
    \caption{Human correlation analysis. We report Spearman's $\rho$ with $p$-values, exact agreement (Agr.), and MAE for human--human agreement (H--H), summary-based scoring (L\textsubscript{s}--H), audio-based direct scoring (L\textsubscript{d}--H), and caption-based scoring (L\textsubscript{c}--H).}
    \label{tab:human_corr}
    \vspace{-2mm}
    \resizebox{\linewidth}{!}{
    \begin{tabular}{ll cccc}
        \toprule
        \textbf{Dim.} & \textbf{Comp.} & \textbf{$\rho\uparrow$} & \textbf{$p$-value} & \textbf{Agr.$\uparrow$} & \textbf{MAE$\downarrow$} \\
        \midrule
        \multirow{4}{*}{Emotion}
            & H--H              & 0.584 & $8.07{\times}10^{-17}$     & 0.438 & 0.586 \\
            & L\textsubscript{s}--H & 0.515 & $2.60{\times}10^{-5}$ & 0.567 & 0.734 \\
            & L\textsubscript{d}--H & 0.044 & 0.73  & 0.500 & 0.994 \\
            & L\textsubscript{c}--H & 0.047 & 0.71  & 0.409 & 1.112 \\
        \midrule
        \multirow{4}{*}{Style}
            & H--H              & 0.514 & $4.91{\times}10^{-12}$     & 0.544 & 0.456 \\
            & L\textsubscript{s}--H & 0.427 & $2.01{\times}10^{-3}$ & 0.640 & 0.568 \\
            & L\textsubscript{d}--H & 0.158 & 0.21  & 0.515 & 0.952 \\
            & L\textsubscript{c}--H & 0.147 & 0.24  & 0.530 & 0.803 \\
        \midrule
        \multirow{4}{*}{NV}
            & H--H              & 0.789 & $5.28{\times}10^{-28}$     & 0.651 & 0.476 \\
            & L\textsubscript{s}--H & 0.518 & $2.66{\times}10^{-4}$ & 0.644 & 0.874 \\
            & L\textsubscript{d}--H & -0.009 & 0.95 & 0.200 & 2.422 \\
            & L\textsubscript{c}--H & 0.193 & 0.20  & 0.511 & 1.207 \\
        \bottomrule
    \end{tabular}
    }
    \vspace{-2mm}
\end{table}

Spearman's $\rho$ measures whether two sets of scores rank samples in a similar order. A lower $p$-value ($< 0.05$) indicates that the correlation is unlikely to arise by chance. In our study, the summary-based scorer shows statistically significant correlation with human judgments on all three dimensions.

For emotion, the summary-based scorer achieves $\rho = 0.515$ ($p < 10^{-4}$). This is close to the human--human correlation of 0.584. Its exact agreement is also high (0.567), although its MAE remains higher than human--human agreement. 
NV preservation shows the strongest human--human correlation ($\rho = 0.789$), which is expected because NVs are less subjective than emotion or scenario style. 
The summary-based scorer reaches $\rho = 0.518$ with high exact agreement (0.644). This supports the use of text-with-NV annotations for NV preservation scoring.
Scenario style is more subjective than NV preservation. Even so, the summary-based scorer reaches significant correlation with human judgments ($\rho = 0.427$, $p < 0.01$), showing that it captures useful style-consistency signals.

Across all dimensions, the summary-based scorer is the only LLM judge design that reaches significant correlation with human judgments. Audio-based direct scoring and caption-based scoring do not show significant correlation in this study. This suggests that a decomposed judging pipeline is more reliable than directly comparing raw audio or long captions at the current stage.

\section{Conclusion}
\label{sec:conclusion}

We introduced \benchmark, a 32.6-hour Chinese--English benchmark that evaluates S2ST systems on translation fidelity and expressiveness preservation (emotion, scenario style, and NVs). A curation pipeline converts real-world audio into structured source annotations with automatic quality control and human validation. \benchmark scores expressiveness through a reference-free LLM-as-a-judge framework that requires no matched target-speech references. Human correlation studies confirm statistically significant agreement with human judgments across expressive dimensions.

Evaluation of six S2ST systems shows that translation accuracy is strong, but expressiveness preservation lags behind. Cascaded systems achieve high BLEU yet score below 1.73 on emotion, and adding expressive instructions to TTS brings almost no improvement. NV preservation benefits from explicit text-level markers (best 2.31/5), whereas end-to-end systems that lack such representation score below 1.58. Duration alignment also requires dedicated control: only the system with explicit duration modeling achieves high SLC scores. Across all findings, the bottleneck in current S2ST lies in expressive transfer rather than semantic transfer. The benchmark data, evaluation prompts, and scoring scripts will be released to support reproducible evaluation of expressive S2ST.

\section*{Limitations}
\begin{itemize}[leftmargin=*, itemsep=1pt, topsep=2pt]
    \item \benchmark currently covers Chinese$\leftrightarrow$English. The pipeline is language-agnostic and extensible to other language pairs, and we plan to expand coverage in future work. 
    \item The expressiveness evaluation relies on an LLM-based caption-then-summarize pipeline, whose accuracy is bounded by the captioning model's ability to perceive paralinguistic cues. As audio-language models improve, the evaluation framework will benefit from stronger perception modules. The scenario style dimension remains the most subjective among the three expressiveness axes, and the current judge still has room for improvement on this dimension.
\end{itemize}

\section*{Ethical Considerations}
The audio data used in \benchmark is collected from publicly accessible web sources. We do not redistribute audio from sources without clear redistribution permission. For such data, we release metadata and annotations only. Samples containing private or sensitive content identified during filtering and human validation are removed. All human annotators involved in quality validation and correlation studies are compensated participants or co-authors, and all provided informed consent for their participation.

\bibliography{ref}

\appendix
% Appendix
\newpage
\section{Details of Pipeline}
\label{app:pipeline}

\subsection{Preprocessing}
\label{app:preprocessing}

This section provides implementation details for the preprocessing stages described in Section~\ref{ssec:collection}. All audio is first converted to 16\,kHz mono WAV format before entering the pipeline.

\paragraph{Source separation.}
We apply BS-Roformer, a MelBand-based audio source separation model using Roformer architecture (distilled checkpoint), to separate foreground speech from background music and environmental noise. Inference runs on a single GPU.

\paragraph{Speaker segmentation.}
Long recordings are first split into manageable chunks to avoid memory overflow. We then apply a joint diarization framework combining three components: (1)~a pyannote segmentation model for detecting speaker change boundaries, (2)~CAM++ speaker verification models for extracting speaker embeddings used in clustering, and (3)~Silero VAD for precise speech activity detection. Adjacent segments from the same speaker are merged, and the final output retains only utterances between 3 and 30 seconds.

\paragraph{Quality filtering.}
Quality filtering proceeds in four stages:
\begin{enumerate}[leftmargin=*, itemsep=1pt, topsep=2pt]
    \item \textit{Bandwidth verification}: We compute the STFT and analyze spectral energy distribution across frequency bands. Audio with an effective bandwidth below 8\,kHz (i.e., narrowband or telephone-quality) is discarded.
    \item \textit{SNR and reverberation}: We estimate per-frame signal-to-noise ratio (SNR) and clarity index (C50, in dB) using the pyannote/brouhaha model, then compute segment-level statistics. We require both metrics to satisfy: 25th-percentile $\geq$ 25\,dB and mean $\geq$ 45\,dB.
    \item \textit{DNSMOS}: We score each segment using DNSMOS (Deep Noise Suppression Mean Opinion Score), a neural perceptual quality predictor. Segments with an overall MOS below 3.0 (on a 5-point scale) are removed.
\end{enumerate}

\paragraph{Language identification.}
We run Whisper-large-v3-turbo on each segment for language identification. Only segments identified as Chinese or English are retained.

\subsection{Translation}
\label{app:translation}

We employ Qwen3-30B-A3B as the translation engine. The prompts are as follows:

\begin{tcolorbox}[colback=gray!5, colframe=black!40, boxrule=0.4pt, arc=2pt, title={\small Translation Prompt (Chinese $\rightarrow$ English)}]
{\small\ttfamily Translate the original text into English, outputting only the translation without adding any explanations or guiding words. Translate the following text into English:\textbackslash n\\
\{text\}}
\end{tcolorbox}

% \begin{tcolorbox}[colback=gray!5, colframe=black!40, boxrule=0.4pt, arc=2pt, title={\small Translation Prompt (English $\rightarrow$ Chinese)}]
% {\small\ttfamily 将原文内容翻译到简体中文，只输出译文，不要添加任何说明或引导词。将下面的文本翻译成中文：\textbackslash n\\
% \{text\}}
% \end{tcolorbox}

\noindent The English-Chinese prompt is the Chinese equivalent of English version.

\section{Expressiveness Annotation Prompts}
\label{app:annotation_prompts}

This section provides the exact prompts used in the two-stage \emph{caption-then-summarize} annotation pipeline described in Section~\ref{ssec:annotation}.

\subsection{Stage 1: Audio Captioning}
\label{app:caption_prompt}

Qwen3-Omni-Captioner receives each audio utterance together without prompt. 

\subsection{Stage 2: Emotion and Style Summarization}
\label{app:summary_prompt}

Given the verbose caption produced by Stage~1, a text LLM (Qwen3-30B-A3B) summarizes it into structured emotion and style fields using the following prompt:

\begin{tcolorbox}[colback=gray!5, colframe=black!40, boxrule=0.4pt, arc=2pt, title={\small Emotion/Style Summarization Prompt}]
{\small\ttfamily You are an expert audio analyst specializing in paralinguistics and speech-scene classification.\\[4pt]
Your task is to analyze a raw audio caption and extract the \textbf{Emotion} and \textbf{Style} needed by a human-aligned speech similarity rubric.\\[4pt]
\textbf{Input Analysis Rules:}\\
1. \textbf{Ignore the Transcription}: Do not judge emotion or style only from what is said. Use vocal delivery, production, and scene cues in the caption.\\
2. \textbf{Emotion}: First infer a concise emotional state such as neutral, cheerful, sad, angry, anxious, excited, serious, or playful. Describe the vocal emotion category, intensity, and subtle attitude/nuance.}
\end{tcolorbox}

\begin{tcolorbox}[colback=gray!5, colframe=black!40, boxrule=0.4pt, arc=2pt, title={\small Emotion/Style Summarization Prompt}]
{\small\ttfamily
3. \textbf{Style}: Here ``style'' means likely scene or media genre, such as film/TV drama, interview, news, audiobook, advertisement, explainer, online class, speech, or livestream. Describe the scene/genre as a short, natural sentence.\\
4. \textbf{No acoustic-only emotion}: The emotion field should describe the perceived feeling or attitude, not only physical traits such as pitch or speed.\\ [4pt]
\textbf{Output Requirements:}\\
- Output strictly in JSON format.\\
- The values for ``emotion'' and ``style'' must be short, descriptive, natural sentences.\\[4pt]
\textbf{Input Caption:}\\
"""\{caption\}"""\\[4pt]
\textbf{JSON Schema:}\\
\{\\
\quad "emotion": "A short sentence describing vocal emotion category, intensity, and attitude.",\\
\quad "style": "A short sentence describing the likely scene or media genre."\\
\}}
\end{tcolorbox}

\section{LLM Judge Prompts}
\label{app:judge_prompts}

This section presents the system prompts used by the LLM-based expressiveness judges described in Section~\ref{ssec:judge}. Each judge receives a system prompt defining the evaluation rubric, followed by a user message containing the source transcript and paired reference/hypothesis descriptions.

\paragraph{Implementation details.}
The text-based LLM judge uses Qwen3-30B-A3B served via vLLM with default decoding parameters (temperature and top-$p$ are not overridden) and a maximum generation length of 2{,}048 tokens. Each sample is scored three times independently under these settings, and the final score is obtained via the aggregation rule described in Appendix~\ref{app:aggregation}.

\subsection{Emotion Judge}
\label{app:emotion_judge}

\begin{tcolorbox}[colback=gray!5, colframe=black!40, boxrule=0.4pt, arc=2pt, title={\small Emotion Judge System Prompt}]
{\small\ttfamily You are an expert evaluator for speech emotion similarity in speech-to-speech translation systems.\\[4pt]
Your task: Given the original transcript only as optional context, compare the reference emotion description with the hypothesis emotion description and score how consistent the two audio clips sound emotionally.\\[4pt]
Critical rule: judge only vocal expression, not semantic content. First infer a concise emotion label for each side, such as neutral, cheerful, sad, angry, anxious, excited, serious, or playful. Then compare three dimensions: emotion category, intensity, and subtle attitude/nuance.\\[4pt]
\textbf{Scoring Rubric (1--5)}\\
\textbf{5 --- Essentially identical emotion}: Core emotion, intensity, and subtle attitude are all basically consistent.\\
\textbf{4 --- Same emotion with minor differences}: The core emotion is consistent, but intensity or details differ slightly. It can still basically be considered the same emotion.\\
\textbf{3 --- Uncertain / same general direction}: It is not very clear whether they are the same. They are generally in the same direction, but the specific emotion or intensity is clearly different.\\
\textbf{2 --- Basically different emotion}: They are basically not the same emotion, with only coarse similarity.\\
\textbf{1 --- Opposite or completely different emotion}: The emotions are opposite or unrelated.\\
\textbf{Required reasoning}\\[4pt]
- Mention the inferred reference and hypothesis emotion labels.\\
- Explain whether category, intensity, and subtle attitude match.\\
- Do not reward a match merely because the transcript topic implies the same emotion.\\[4pt]
% \end{tcolorbox}

% \begin{tcolorbox}[colback=gray!5, colframe=black!40, boxrule=0.4pt, arc=2pt, title={\small Emotion Judge System Prompt}]
% {\small\ttfamily 
\textbf{Output format}\\
Return a JSON object: \{"score": <int 1-5>, "reason": "<brief explanation>"\}\\
Return ONLY the JSON object, no other text.}
\end{tcolorbox}

\noindent The user message follows the format:
% Transcript: \{ref\_text\}\\
% Reference emotion: \{ref\_emotion\}\\
% Hypothesis emotion: \{hyp\_emotion\}

\begin{tcolorbox}[colback=gray!5, colframe=black!40, boxrule=0.4pt, arc=2pt]
{\small\ttfamily Transcript: \{ref\_text\}\\
Reference emotion: \{ref\_emotion\}\\
Hypothesis emotion: \{hyp\_emotion\}}
\end{tcolorbox}

\subsection{Scenario Style Judge}
\label{app:style_judge}

\begin{tcolorbox}[colback=gray!5, colframe=black!40, boxrule=0.4pt, arc=2pt, title={\small Scenario Style Judge System Prompt}]
{\small\ttfamily You are an expert evaluator for speech style similarity in speech-to-speech translation systems.\\[4pt]
Your task: Given the original transcript only as optional context, compare the reference style description with the hypothesis style description and score how consistent the two audio clips sound in scene/genre.\\[4pt]
Critical rule: judge the speaking scene/genre from vocal delivery and production cues, not the text topic. First summarize each side with a short noun phrase, such as news broadcast, interview, film/TV drama, audiobook, advertisement, online class, explainer, public speech, or livestream. Then compare scene/genre and confidence.\\
\textbf{Scoring Rubric (1--5)}\\
\textbf{5 --- Fully confident same scene/genre}: You are completely confident the scene or genre is consistent.\\
\textbf{4 --- Basically same scene/genre}: It can basically be considered the same scene or genre, with only minor uncertainty or framing differences.\\
\textbf{3 --- Uncertain / same broad category}: It is not very clear whether they are the same. They are in the same broad category but the specific scene differs.\\
\textbf{2 --- Basically different scene/genre}: They are basically not the same scene, but have some similarity, such as both being formal contexts.\\
\textbf{1 --- Completely different}: The scene or genre is completely different.\\
\textbf{Required reasoning}\\
- Mention the inferred reference and hypothesis scene/genre noun phrases.\\
- Explain confidence and whether the specific scene/genre matches.\\
- Treat formality and delivery manner only as supporting cues.\\[4pt]
\textbf{Output format}\\
Return a JSON object: \{"score": <int 1-5>, "reason": "<brief explanation>"\}\\
Return ONLY the JSON object, no other text.}
\end{tcolorbox}

\noindent The user message follows the format:
% Transcript: \{ref\_text\}\\
% Reference style: \{ref\_style\}\\
% Hypothesis style: \{hyp\_style\}
\begin{tcolorbox}[colback=gray!5, colframe=black!40, boxrule=0.4pt, arc=2pt]
{\small\ttfamily Transcript: \{ref\_text\}\\
Reference style: \{ref\_style\}\\
Hypothesis style: \{hyp\_style\}}
\end{tcolorbox}

\subsection{NV Judge}
\label{app:event_judge}

\begin{tcolorbox}[colback=gray!5, colframe=black!40, boxrule=0.4pt, arc=2pt, title={\small NV Judge System Prompt}]
{\small\ttfamily You are an expert evaluator for sound event similarity in speech-to-speech translation systems.\\[4pt]
Your task: Compare the reference text/audio description with embedded sound event tags against the hypothesis text/audio description, and score whether all sound events are preserved after translation. Sound events include breathing, coughing, laughter, sneezing, crying, whispering, sighing, panting, burping, and similar audible events.\\[4pt]
Judge event type, count, order, and approximate position. Also penalize obvious newly added events that were not in the reference, especially salient events such as laughter or coughing.\\[4pt]
\textbf{Scoring Rubric (1--5)}\\[2pt]
\textbf{5 --- All events preserved}: All sound events are preserved, including type, count, order, and approximate position. There are no obvious added events.\\[2pt]
\textbf{4 --- Main and minor events basically preserved}: Major and minor events are basically preserved, with only slight position or intensity deviations. If the only addition is one natural breath, score 4.\\[2pt]
\textbf{3 --- Main events preserved with noticeable errors}: Main events are preserved, but there are clear omissions, position deviations, or several added minor natural breaths that do not dominate the listening impression.\\[2pt]
\textbf{2 --- Most events not preserved / disruptive additions}: Most events are not preserved, event types are clearly wrong, positions are very shifted, or the hypothesis adds salient events not in the reference that interfere with judgment.}
\end{tcolorbox}

\begin{tcolorbox}[colback=gray!5, colframe=black!40, boxrule=0.4pt, arc=2pt, title={\small NV Judge System Prompt}]
{\small\ttfamily
\textbf{1 --- Events clearly not preserved}: It is clear that the source audio's sound events were not preserved.\\[4pt]
\textbf{Required reasoning}\\
- Identify reference and hypothesis events when possible.\\
- Discuss type, count, order, approximate position, and added events.\\[4pt]
\textbf{Output format}\\
Return a JSON object: \{"score": <int 1-5>, "reason": "<brief explanation>"\}\\
Return ONLY the JSON object, no other text.}
\end{tcolorbox}

\noindent The user message follows the format:
\begin{tcolorbox}[colback=gray!5, colframe=black!40, boxrule=0.4pt, arc=2pt]
{\small\ttfamily Reference (text with events): \{ref\_events\}\\
Hypothesis (text with events): \{hyp\_events\}}
\end{tcolorbox}

\section{Auto Quality Assurance Prompt}
\label{app:gemini_prompt}

During quality assurance (Section~\ref{ssec:quality}), a SOTA multimodal LLM receives each audio sample together with its associated annotations and scores them on a 1--5 scale across multiple dimensions. The model listens to the audio directly via multimodal input and evaluates whether the text annotations accurately describe what it hears.

\begin{tcolorbox}[colback=gray!5, colframe=black!40, boxrule=0.4pt, arc=2pt, title={\small Multimodal Scoring Prompt}]
{\small\ttfamily You are a professional speech/audio quality evaluation expert. Please listen carefully to the provided audio and score how accurate the following text information is.\\[4pt]
\#\# Text Information Associated with the Audio\\[2pt]
\textbf{Original Speech Transcript}: \{punctuated\_text\}\\
\textbf{English Translation}: \{translation\_en\}\\
\textbf{Chinese Translation}: \{translation\_zh\}\\
\textbf{Emotion Label}: \{emotion\}\\
\textbf{Style Label}: \{style\}\\
\textbf{Text with Sound Event Tags}: \{text\_with\_events\}\\[4pt]}
\end{tcolorbox}

\begin{tcolorbox}[colback=gray!5, colframe=black!40, boxrule=0.4pt, arc=2pt, title={\small Multimodal Scoring Prompt}]
{\small\ttfamily
\#\# Scoring Task\\[2pt]
Please listen carefully and score the following dimensions (1--5). For any dimension, if the score is less than 5, please provide the \textbf{CORRECT content} in the reason field. If providing the correction is not applicable, explain the error.\\[4pt]
\#\#\# 1. Transcript Accuracy (transcript\_score)\\
- 5: Completely accurate, word-for-word match\\
- 4: Mostly accurate with minor punctuation or spelling errors\\
- 3: Generally correct but with some word errors\\
- 2: Many errors affecting meaning\\
- 1: Completely wrong or missing\\[4pt]
\#\#\# 2. Translation Quality (translation\_score)\\
- 5: Completely accurate, natural and fluent; perfectly preserves the original meaning\\
- 4: Mostly accurate with minor unnatural phrasing or small errors\\
- 3: Generally correct, but with some obvious errors or omissions\\
- 2: Many errors that affect understanding\\
- 1: Severely wrong or completely irrelevant\\[4pt]
\#\#\# 3. Emotion Consistency (emotion\_score)\\
- 5: Fully accurate; highly consistent with the audio emotion\\
- 4: Mostly accurate; captures the main emotion\\
- 3: Partially accurate; some omissions or mild deviations\\
- 2: Clearly inconsistent with the actual emotion\\
- 1: Completely wrong\\[4pt]
\#\#\# 4. Style Consistency (style\_score)\\
- 5: Fully accurate; precisely describes the speaking manner\\
- 4: Mostly accurate\\
- 3: Partially correct\\
- 2: Largely deviates\\
- 1: Completely wrong\\[4pt]}
\end{tcolorbox}

\begin{tcolorbox}[colback=gray!5, colframe=black!40, boxrule=0.4pt, arc=2pt, title={\small Multimodal Scoring Prompt}]
{\small\ttfamily
\#\#\# 5. Sound Event Retention (event\_score)\\
- 5: All sound events are correctly marked and positioned; no omissions or false tags\\
- 4: Most sound events are correctly marked/positioned with minor omissions\\
- 3: Major sound events are marked, but there are some omissions, incorrect tags, or misplaced events\\
- 2: Many incorrect tags, omissions, or wrong positions\\
- 1: Severely wrong or completely missing\\[4pt]
---\\[4pt]
\#\# Output Format\\
Please output the scoring result in JSON format. Only include fields for dimensions that are applicable.}
\end{tcolorbox}

\section{Human Evaluation Criteria}
\label{app:human_eval}

This section details the human annotation criteria used in two evaluation stages: (1)~benchmark quality validation during data curation, and (2)~LLM judge--human alignment evaluation. All annotators are proficient in both English and Chinese, and are experts in speech annotation.

\subsection{Benchmark Quality Validation}
\label{app:benchmark_human_check}

During the final quality assurance stage (Section~\ref{ssec:quality}), human annotators verify the accuracy of automatically generated annotations. Each sample is presented with its audio and associated text metadata. Annotators score on a 1--5 scale across the following dimensions:

\paragraph{Transcript accuracy:}
\begin{itemize}[leftmargin=*, itemsep=1pt]
    \item \textbf{5}: The text perfectly matches the audio with no errors or omissions.
    \item \textbf{4}: Mostly accurate with only 1--2 minor errors (homophones, punctuation) that do not affect comprehension.
    \item \textbf{3}: Generally understandable but with 3--5 errors or missing phrases that slightly affect comprehension.
    \item \textbf{2}: Contains obvious errors or omissions; parts of the content are difficult to understand.
    \item \textbf{1}: Extensive errors or large missing segments; content is severely distorted.
\end{itemize}

\paragraph{Translation quality:}
\begin{itemize}[leftmargin=*, itemsep=1pt]
    \item \textbf{5}: Translation is accurate and fluent, fully conveys the original meaning, and uses natural idiomatic language with no mistranslation or omission.
    \item \textbf{4}: Generally accurate; occasional phrasing is slightly stiff or unnatural but does not affect comprehension.
    \item \textbf{3}: Conveys the main meaning but with partial content deviations or omissions; requires inference to understand.
    \item \textbf{2}: Contains obvious mistranslations or large omissions; overall meaning is distorted.
    \item \textbf{1}: Translation is severely wrong or almost entirely unrelated to the source.
\end{itemize}

\paragraph{Emotion label accuracy:}
\begin{itemize}[leftmargin=*, itemsep=1pt]
    \item \textbf{5}: The emotion label perfectly matches the audio; description is precise (\eg anger, joy, sadness).
    \item \textbf{4}: Generally correct; the emotion category is accurate but the description is slightly overstated or understated.
    \item \textbf{3}: Emotion direction is approximately correct (\eg positive vs.\ negative), but the specific emotion category has deviation.
    \item \textbf{2}: The emotion label does not match the actual audio; obvious confusion (\eg labeling sadness as anger).
    \item \textbf{1}: The emotion label is completely wrong or entirely unrelated to the audio.
\end{itemize}

\paragraph{Scenario style label accuracy:}
\begin{itemize}[leftmargin=*, itemsep=1pt]
    \item \textbf{5}: The style description perfectly matches the speaking manner (\eg storytelling, debate, reading aloud).
    \item \textbf{4}: Style is generally correct but the description is overly broad or lacks detail.
    \item \textbf{3}: The style judgment has some basis but there is obvious deviation or confusion between different styles.
    \item \textbf{2}: The style description basically does not match the actual speaking manner.
    \item \textbf{1}: The style description is completely wrong and unrelated to the audio characteristics.
\end{itemize}

\paragraph{NVs accuracy:}
\begin{itemize}[leftmargin=*, itemsep=1pt]
    \item \textbf{5}: All NV events (including type and position) are correctly annotated with no false positives or omissions.
    \item \textbf{4}: Event types are correct with slight position deviations; at most 1 minor event is missed or falsely added.
    \item \textbf{3}: Major events are identified but 1--2 event types are confused (\eg labeling a sigh as breathing).
    \item \textbf{2}: Event annotations are chaotic with multiple false positives or missed salient events.
    \item \textbf{1}: Most events are fabricated or omitted; annotations severely disagree with the actual audio.
\end{itemize}

\paragraph{Audio quality:}
\begin{itemize}[leftmargin=*, itemsep=1pt]
    \item \textbf{5}: Audio is very natural and clear with no noise, distortion, or anomalies.
    \item \textbf{4}: Mostly natural and clear with very slight noise or unnaturalness that does not affect comprehension.
    \item \textbf{3}: Acceptable; some noise or unnaturalness exists but content remains intelligible.
    \item \textbf{2}: Poor audio quality with obvious noise or unnatural sound that affects comprehension.
    \item \textbf{1}: Severely distorted, extremely noisy, or completely unnatural audio that is difficult to discern. Score~1 for: (i)~more than one speaker present, (ii)~language mismatch with pool classification, or (iii)~prominent background music or singing.
\end{itemize}

For efficiency, part of the quality validation in normal subset is conducted using a binary accept/reject protocol rather than the full 1--5 rating scheme. Annotators are asked to reject samples with clear issues and to accept samples that satisfied the benchmark quality requirements. The 1--5 criteria described above were used to define the validation standards and to guide annotators' decisions.

\subsection{LLM Judge--Human Alignment Evaluation}
\label{app:judge_human_eval}

To validate LLM judge reliability, human annotators independently score the same samples on which the judge operates. Annotators listen to both the source audio and the system-translated hypothesis audio, then rate expressiveness preservation. The scoring criteria are as follows.

\paragraph{Emotion similarity.}
Annotators are instructed: ``How consistent is the emotion between the two audio clips? Higher similarity yields a higher score.'' Guidelines specify:
\begin{itemize}[leftmargin=*, itemsep=1pt]
    \item Judge only vocal expression, not text content.
    \item First summarize the emotion of each clip with a single word (\eg neutral, cheerful, sad, angry, anxious, excited, serious, playful).
    \item Then compare three dimensions: emotion category, intensity, and subtle attitude.
\end{itemize}
The rubric is:
\begin{itemize}[leftmargin=*, itemsep=1pt]
    \item \textbf{5}: Core emotion, intensity, and subtle attitude are all basically consistent.
    \item \textbf{4}: Core emotion is consistent; intensity or detail differs slightly. Can still be considered the same emotion.
    \item \textbf{3}: Not clearly the same; generally in the same direction, but specific emotion or intensity differs noticeably (\eg both are positive, but one is very excited while the other is only lightly upbeat).
    \item \textbf{2}: Basically not the same emotion; only very coarse similarity remains (\eg both are not happy, or both are relatively calm, but actual impression differs substantially).
    \item \textbf{1}: Emotions are opposite or completely different (\eg sadness vs.\ happiness, calm narration vs.\ excited performance).
\end{itemize}

\paragraph{Scenario Style similarity.}
Annotators are instructed: ``Here `style' refers to the scene/genre where the audio might occur, such as film/TV drama, interview, news, audiobook, advertisement, commentary, etc. How consistent is the style between the two audio clips?'' Guidelines specify:
\begin{itemize}[leftmargin=*, itemsep=1pt]
    \item First summarize each clip with a short noun phrase: news broadcast, interview, film/TV drama, audiobook, advertisement, online class, commentary, customer service, speech, livestream, etc.
    \item Then compare: scene/genre match and confidence level.
\end{itemize}
The rubric is:
\begin{itemize}[leftmargin=*, itemsep=1pt]
    \item \textbf{5}: Fully confident that the scene/genre is the same.
    \item \textbf{4}: Can basically be considered the same scene/genre.
    \item \textbf{3}: Not clearly the same; same broad category but specific scene differs (\eg both sound like performed speech, but one is film/TV drama while the other is audiobook).
    \item \textbf{2}: Can basically be considered different scenes, but with some shared property (\eg both are formal settings).
    \item \textbf{1}: Completely different (\eg audiobook performance vs.\ advertisement voice-over).
\end{itemize}

\paragraph{NVs similarity.}
Annotators are instructed: ``Are all sound events preserved between the source and translated audio? Sound events include breathing, coughing, laughter, etc.'' The rubric is:
\begin{itemize}[leftmargin=*, itemsep=1pt]
    \item \textbf{5}: All NV events are preserved in type, count, order, and approximate position; no obvious added events.
    \item \textbf{4}: Main and minor NV events are mostly preserved with only slight position or intensity deviations; adding one natural breath is acceptable.
    \item \textbf{3}: Main NV events are preserved but with clear omissions, position deviations, or several added minor breaths that do not dominate the listening impression.
    \item \textbf{2}: Most NV events are not preserved, types are clearly wrong, positions are very shifted, or salient events (especially laughter or coughing) are added that were not present in the source.
    \item \textbf{1}: Source audio's NV events are clearly not preserved.
\end{itemize}

\paragraph{Inter-annotator consistency filtering.}
We apply simple inter-annotator quality control: samples where annotator scores on the target dimension show substantial divergence are excluded, and individual outlier annotations that deviate from the remaining scores are removed. The final human reference score is the mean of the retained annotations.

\subsection{Consent and Data Use}
We obtained permission to use the collected annotation data for research evaluation. The permission covers data curation, annotation, and use in the benchmark under the intended research setting. Records of the authorization are maintained internally.

All annotators consented to the use of their annotations for academic research. They were compensated at market-competitive rates through the corresponding annotation workflow. Annotators were well educated, fluent in both Chinese and English, and had prior experience with audio annotation tasks.

\section{Automatic Score Aggregation}
\label{app:aggregation}

Each expressiveness dimension is scored three times independently per sample. We apply the following aggregation rule to obtain the final score:
\begin{itemize}[leftmargin=*, itemsep=1pt, topsep=2pt]
    \item If all three scores are identical, that score is used directly.
    \item If two scores are identical and the third differs by at least 2 points, the third is treated as an outlier and removed. The final score is the agreed value.
    \item If two scores are identical and the third differs by less than 2 points, the final score is the mean of all three.
    \item If all three scores are different, the final score is the median.
\end{itemize}
These rules are applied to the automatic scores and do not introduce any bias.

\section{Evaluated System Details}
\label{app:system_details}

\paragraph{Three-Stage pipeline.}
The Three-Stage system operates as follows. First, source speech is transcribed using Qwen3-ASR~\citep{shi2026qwen3asrtechnicalreport}, and NVs are detected using the same BEATs-based NV detection module as used in \benchmark annotation. The resulting transcription and NV markers are passed to Qwen3-235B-A22B~\citep{yang2025qwen3technicalreport}, which produces the target-language translation along with an instruction describing the inferred emotion and scenario style. For the NV subset, the LLM additionally performs translation with inline event tag generation (\eg translating ``\texttt{[laughter] that's funny}'' with NV markers preserved in the output). VoxCPM2~\citep{voxcpm2_2026} is the TTS module, receiving the translated text, a reference audio for voice cloning, and the style instruction. The \emph{w/o Instruct} variant removes the style instruction, providing only text and reference audio to TTS.

\paragraph{Two-Stage pipeline.}
The Two-Stage system uses Qwen3-Omni-Instruct~\citep{xu2025qwen3omnitechnicalreport} as an audio-input multimodal LLM that performs direct speech-to-text translation without explicit ASR or NV detection. In addition to translating, Qwen3-Omni-Instruct infers emotion and scenario style from the input speech to generate a style instruction for TTS. The translated text, reference audio, and style instruction are then passed to VoxCPM2 for voice-cloning TTS synthesis. The \emph{w/o Instruct} variant removes the style instruction.

\section{Experiments}
\label{app:results}

\subsection{Metric Definitions}
\label{app:metrics}

We describe all evaluation metrics used in \benchmark. For all metrics, higher values indicate better performance.

\begin{itemize}[leftmargin=*, itemsep=2pt]

\item \textbf{BLEU} evaluates translation fidelity by computing corpus-level BLEU scores using the SacreBLEU library. We apply language-specific preprocessing: English text is lowercased and stripped of punctuation (excluding apostrophes), while Chinese text is normalized to simplified characters, punctuation is removed, and characters are separated by spaces. Chinese samples use the \texttt{zh} tokenizer; English samples use the \texttt{13a} tokenizer. For speech-output evaluation, the generated speech is first transcribed with Qwen3-ASR~\citep{shi2026qwen3asrtechnicalreport} before computing BLEU against the reference translation.

\item \textbf{COMET} measures translation quality using neural learned metrics. We use the \texttt{wmt22-comet-da} model for reference-based scoring, which takes source text, hypothesis translation, and reference translation as input and produces a quality score.

\item \textbf{XCOMET} provides both reference-based and quality-estimation (QE) translation scores using the \texttt{XCOMET-XL} model. The reference-based variant uses source, hypothesis, and reference; the QE variant uses only source and hypothesis, requiring no reference translation. Both variants produce scores on a 0--1 scale.

\item \textbf{Speaker Similarity (SIM)} measures voice preservation as the cosine similarity between speaker embeddings extracted from source and translated speech. We use the WavLM-Large + ECAPA-TDNN speaker verification model from UniSpeech~\citep{Wang2021UniSpeech}, following the same pair-by-pair inference protocol as Seed-TTS-eval\footnote{https://github.com/BytedanceSpeech/seed-tts-eval}. Higher values indicating greater speaker similarity.

\item \textbf{SLC-0.2 and SLC-0.4} (Speech Length Compliance) assess duration alignment between source and translated speech. SLC measures the proportion of utterances whose output-to-input duration ratio falls within $[1-\epsilon, 1+\epsilon]$, where $\epsilon \in \{0.2, 0.4\}$. These metrics are relevant to video dubbing and simultaneous interpretation, where the translated speech must match the timing of the source.

\item \textbf{Expressiveness scores} (Emotion, Style, NV) are produced by the LLM-as-a-judge framework described in Section~\ref{ssec:judge}, with scoring rubrics detailed in Appendix~\ref{app:judge_prompts}. Each dimension is scored on a 1--5 scale, aggregated over three independent runs as described in Appendix~\ref{app:aggregation}.

\end{itemize}

\subsection{More Discussions}
\paragraph{Translation fidelity of cascaded systems}
We observe that the three-stage system obtains lower BLEU than the two-stage system. After manual inspection, we find that the larger model used in the three-stage pipeline is more likely to generate content unrelated to the translation, such as meta-instructions like ``let me translate this for you.'' Such hallucinated or extraneous text lowers translation fidelity.

\paragraph{Low voice SIM score}
Voice SIM scores are generally low, especially compared with voice similarity scores commonly reported in TTS. This is because the reference and target speech are in different languages in S2ST evaluation. Cross-lingual differences can reduce speaker similarity scores. This issue is common in S2ST evaluation.

\section{Frequently Asked Questions}
\label{app:faq}

\subsection{Why not use classifiers for expressiveness evaluation?}
A natural alternative to our LLM-as-a-judge framework is to classify the source and translated speech independently, and then map each pair of predicted labels to a preservation score. For example, one could assign a high score when the source and target share the same emotion label, and a lower score when their labels differ.

We do not adopt this strategy as the main evaluator because expressive preservation is not only a label-matching problem. The same source--target label pair may correspond to different preservation quality depending on expression intensity, mixed emotions, discourse context, and how the expression is realized in speech. Scenario style also depends on lexical content, speaker role, and delivery intent, rather than purely on acoustic cues. NV preservation further requires event-level comparison, including whether an NV is preserved, omitted, newly inserted, or placed at an inappropriate position. Moreover, we aim to represent speech with richer textual descriptions rather than a small set of predefined labels, allowing the evaluator to capture fine-grained expressive cues beyond closed-set categories.

Therefore, \benchmark formulates expressive S2ST evaluation as a contextual, comparative, and graded scoring problem. The LLM judge compares structured source and hypothesis annotations under a shared rubric and assigns 1--5 preservation scores. We validate these automatic scores against human judgments in Section~\ref{ssec:human_correlation}.

\subsection{Why not extend to more languages?}

Currently, \benchmark covers only Chinese and English. We view this work as an initial study of the data curation pipeline and reference-free evaluation methodology for expressive S2ST. The proposed pipeline is not tied to a specific language pair, and we believe it can be extended to other high-resource languages with available ASR, translation, alignment, and speech-understanding tools. Extending \benchmark to more languages, especially low-resource settings, is an important direction for future work.

\end{document}